\documentclass[conference]{IEEEtran}
\IEEEoverridecommandlockouts
\usepackage{cite}
\usepackage{amsmath,amssymb,amsfonts}
\usepackage{algorithmic}
\usepackage{graphicx}
\usepackage{textcomp}
\usepackage{xcolor}
\usepackage{booktabs}
\usepackage{pgfplots}
\usepackage{tikz}
\pgfplotsset{compat=1.18}

\usepackage{listings}
\usepackage{algorithm}
\usepackage{amsmath}
\usepackage{tcolorbox}
\usepackage{multirow}
\usepackage{amssymb}
\usepackage{threeparttable} 
\usepackage{caption}
\usepackage{enumitem}

\def\BibTeX{{\rm B\kern-.05em{\sc i\kern-.025em b}\kern-.08em
    T\kern-.1667em\lower.7ex\hbox{E}\kern-.125emX}}

\begin{document}

\title{Blueprint First, Model Second: A Framework for Deterministic LLM Workflow
\thanks{\textsuperscript{\textdagger}~Corresponding author: Yuhang Ye (yeyu.yyh@alibaba-inc.com).}
}

\author{
\IEEEauthorblockN{Libin Qiu, Yuhang Ye\textsuperscript{\textdagger}, Zhirong Gao, Xide Zou,\\
Junfu Chen, Ziming Gui, Weizhi Huang,\\
Xiaobo Xue, Wenkai Qiu, Kun Zhao}
\IEEEauthorblockA{\textit{Alibaba Group}, Hangzhou, China\\
\{libin.qlb, yeyu.yyh, gaozhirong.gzr, zouxide.zxd, chenjunfu.cjf,\\
guiziming.gzm, qingli.hwz, xiaobo.xxb, yiosng.qwk, zhaokun.zk\}@alibaba-inc.com}
}

\maketitle

\begin{abstract}
While powerful, the inherent non-determinism of large language model (LLM) agents limits their application in structured operational environments where procedural fidelity and predictable execution are strict requirements. This limitation stems from current architectures that conflate probabilistic, high-level planning with low-level action execution within a single generative process. To address this, we introduce the \textsc{Source Code Agent} framework, a new paradigm built on the ``Blueprint First, Model Second'' philosophy that decouples workflow logic from the generative model. An expert-defined operational procedure is first codified into a source code-based Execution Blueprint, which is then executed by a deterministic engine. The LLM is strategically invoked as a specialized tool to handle bounded, complex sub-tasks within the workflow, but never to decide the workflow's path. We evaluate on the TravelPlanner benchmark for constraint-aware travel planning. The \textsc{Source Code Agent} achieves a 35.56\% final pass rate, a 97.6\% improvement over the state-of-the-art ATLAS baseline (18.00\%) on the same Claude-Sonnet-4 backbone. Critically, it reduces constraint violations by 96.0\% (11 vs 275) while improving execution efficiency by 27.1\% (10.2$\pm$0.7 steps vs 14.0). Two production incident-diagnosis deployments and additional results on ScienceWorld and ALFWorld confirm that the architecture transfers beyond travel planning to procedurally well-defined, constraint-intensive workflows. Our work enables the verifiable and reliable deployment of autonomous agents in applications governed by strict procedural logic.
\end{abstract}

\begin{IEEEkeywords}
LLM Agents, Deterministic Workflow, Agent Reliability
\end{IEEEkeywords}

\section{Introduction}
\label{sec:intro}
With dramatic evolution in recent years, Foundation Models (FMs), such as the GPT~\cite{openai2023gpt4} and Claude~\cite{claude} model families, are increasingly being adopted as the core reasoning engine for powerful, general-purpose agents capable of tackling complex tasks~\cite{wang2024survey}. To solve real-world problems, these agents are typically architected as compound systems integrating essential components for planning, memory, and tool use~\cite{liu2024graphcoder,zhao2024repair}. While these capabilities grant agents unprecedented flexibility for creative and exploratory endeavors, this flexibility introduces a critical challenge for a different class of tasks: the inherent non-determinism in their execution trajectories~\cite{zhang2025llm}. For applications where reliability, safety, and predictability are non-negotiable, this operational uncertainty poses an insurmountable barrier to adoption~\cite{shi2025know}.

This limitation becomes starkly evident in high-stakes, safety-critical domains. Consider the task of enterprise software troubleshooting, such as resolving a \texttt{Java OutOfMemoryError} in a production environment. An experienced human engineer follows a deterministic, highly optimized workflow: check garbage collection statistics using \texttt{jstat}, generate a heap dump with \texttt{jmap} if the old generation is full, and then analyze the dump file to pinpoint the memory leak. In contrast, a general-purpose agent, guided by its end-to-end reasoning, might issue broad, exploratory queries or execute a sequence of diagnostic commands that, while potentially correct, is unpredictable and significantly slower. This lack of a guaranteed execution path renders such agents unsuitable for tasks where every step must be precise, verifiable, and adhere to established operational protocols. 

\begin{figure}
    \centering
    \includegraphics[width=0.98\linewidth]{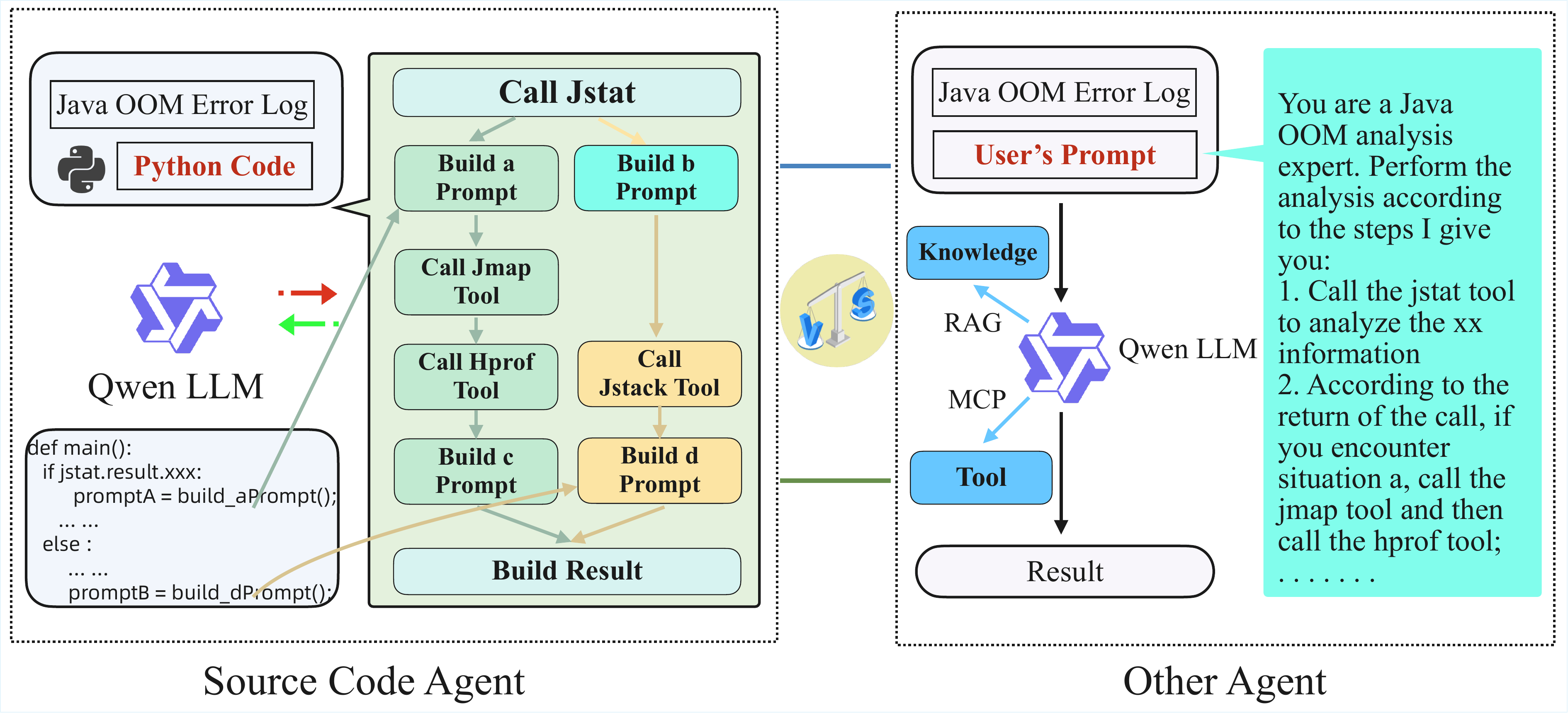}
    \caption{\textsc{Source Code Agent} vs. Other Agent}
    \label{fig:diff}
\end{figure}

The source of this unpredictable behavior is not a flaw in the agent's intelligence, but a byproduct of its core architectural design. Current agent frameworks typically entrust the entirety of the decision-making process—from high-level strategic reasoning to low-level action selection—to the probabilistic outputs of a language model. While this end-to-end, generative approach provides remarkable adaptability for open-ended problems, it inherently conflates the \textit{what} (the strategic goal) with the \textit{how} (the specific execution steps). Consequently, the agent's path is not guided by a strict, predefined logic but is instead regenerated at each step, making guarantees of sequence, correctness, and efficiency difficult to enforce. This creates an architectural gap for tasks demanding procedural fidelity. 

To address this challenge, we introduce a new architectural paradigm, \textbf{``Blueprint First, Model Second''} and present its concrete implementation, the \textbf{\textsc{Source Code Agent}} framework. In our framework, an expert-defined operational procedure is first codified into a machine-readable \textit{Execution Blueprint}. This blueprint is not an abstract diagram but a concrete workflow realized as \textbf{source code} (e.g., a Python script) that explicitly defines the sequence of steps, conditional logic, and decision points. A deterministic engine then executes this code-defined blueprint, navigating its states with complete fidelity. The role of the Foundation Model is thus strategically reframed: it is no longer the central decision-maker but is invoked as a specialized tool at specific nodes of the blueprint to handle complex but bounded sub-tasks, such as parsing an error log or summarizing a command's output. This separation of concerns—where a deterministic engine manages the workflow blueprint and the intelligent model handles discrete task execution—transforms the agent’s behavior from an unpredictable exploration into a verifiable and auditable process. Figure \ref{fig:diff} shows the difference between the source code agent and other agents in handling OOM errors.

We validate the effectiveness of our framework through a comprehensive evaluation on the TravelPlanner benchmark~\cite{xie2024travelplanner}, which is designed for constraint-aware travel planning with complex user requirements and commonsense rules. The results are compelling: the \textsc{Source Code Agent} achieves a 35.56\% final pass rate, representing a 97.6\% improvement over the state-of-the-art ATLAS baseline (18.00\%). Most significantly, our framework reduces constraint violations by 96.0\% (11 total violations vs ATLAS's 275), with zero violations achieved on four constraint types through code-based enforcement. Furthermore, our approach improves execution efficiency by 27.1\%, reducing average steps from 14.0 (ReAct) to 10.2$\pm$0.7. These results validate our framework as a robust solution for deploying agents in constraint-intensive scenarios where procedural adherence and verifiable execution are strict requirements.
\section{Related Work}
\label{sec:related}

\subsection{LLM-Based Agent Architectures}

\paragraph{Planning and Reasoning Agents}
ReAct~\cite{yao2022react} pioneers the interleaving of reasoning traces with tool-using actions, enabling dynamic replanning based on environmental observations. Reflexion~\cite{shinn2023reflexion} extends this paradigm by incorporating verbal self-reflection from failed trajectories, storing episodic experiences to guide future attempts~\cite{wei2022chain,yao2023tree,besta2024graph}. While these methods demonstrate impressive capabilities in open-ended tasks, they rely fundamentally on probabilistic LLM generation for both strategic planning and tactical execution, leading to non-deterministic behavior that limits their applicability in structured operational environments~\cite{schick2023toolformer,patil2024gorilla,arithmetic2024cot}.

\paragraph{Multi-Agent Collaboration}
Recent work explores decomposing complex tasks across specialized agents. ATLAS~\cite{choi2025atlas} employs dedicated agents for search, constraint management, planning, and verification in travel planning tasks, achieving 18\% final pass rate on TravelPlanner benchmark. MetaGPT~\cite{hong2023metagpt} organizes agents into software development roles with structured communication protocols. While multi-agent decomposition improves performance through specialization, these systems still coordinate via natural language prompts, inheriting the uncertainty of LLM-based constraint enforcement. ATLAS, despite its sophisticated architecture, makes 275 constraint violations per 1000 tasks due to the probabilistic nature of prompt-based rule following~\cite{guo2024large,masterman2024landscape}.

\subsection{Code-Based and Structured Reasoning}

\paragraph{Executable Reasoning}
Chain-of-Code~\cite{li2024chain} generates Python-like pseudocode for semantic sub-tasks, using an ``LMulator" to simulate execution when code semantics are ambiguous. Logic-LM~\cite{pan2023logic} translates natural language problems into symbolic formulations, invoking external solvers for faithful logical inference. These approaches leverage code for \textit{problem decomposition} or \textit{verification}, but the LLM remains the primary controller for high-level planning and constraint management. In contrast, our framework inverts this relationship: source code controls the workflow, while the LLM is invoked as a specialized tool for bounded sub-tasks~\cite{jung2025code,xu2024faithful}.

\paragraph{Code as the Action Space}
CodeAct~\cite{wang2024executable} consolidates an agent's actions into executable Python, letting the LLM emit code instead of fixed-schema tool calls and observe the interpreter's feedback. This improves the \textit{execution} layer—how a single action is expressed and grounded—but the LLM still decides, at each turn, which action to take next; the control flow remains a probabilistic generation. Our framework operates at a different layer and is therefore complementary rather than competing. The Source Code Agent fixes the \textit{control flow} as a deterministic, expert-authored program and invokes the model only at bounded nodes. The two compose naturally: an Execution Blueprint can orchestrate the overall path while a CodeAct-style generator handles open-ended synthesis inside an individual node. The distinction matters in industrial settings where a single mis-ordered action (e.g., dumping heap on a production-traffic host, or issuing an order-processing call out of sequence) is costly—a guarantee that turn-by-turn code generation alone cannot provide~\cite{yang2024swe}.

\paragraph{Plan-and-Execute Paradigms}
Plan-and-Solve~\cite{wang2023plan} proposes generating a complete plan upfront before executing tools, reducing interleaving overhead. However, the plan itself is still generated via LLM inference, inheriting uncertainty. Our execution blueprints are fundamentally different: they are not LLM-generated plans but \textit{expert-codified} workflows that encode procedural knowledge as deterministic control flow. This architectural distinction transforms the agent from an autonomous planner into a reliable executor of pre-verified procedures.

\subsection{Positioning of Our Work}

Our Source Code Agent framework occupies a distinct position in the agent architecture landscape:

\textbf{Architectural Control.} Unlike ReAct and Reflexion, where the LLM controls workflow through probabilistic generation, we use deterministic source code as the primary control mechanism. The LLM is strategically relegated to bounded sub-tasks with clear input-output interfaces, eliminating workflow-level uncertainty.

\textbf{Constraint Enforcement.} In contrast to ATLAS's natural language-based Constraint Manager, we encode constraints as programmatic checks. The distinction is fundamental: prompt-based constraints remain uncertain (``ensure no repeated restaurants" can still be violated), while code-based constraints are certain (\texttt{if restaurant.id in used:} \texttt{raise ValidationError()} cannot be bypassed).

\textbf{Separation of Concerns.} While Chain-of-Code and Logic-LM use code for task decomposition or verification, our framework uses source code as the \textit{primary workflow definition}. This shift from ``LLM with code tools" to ``code with LLM tool" represents a paradigm change in how agents achieve reliability—moving from adaptive uncertainty management to architectural uncertainty elimination.

The key insight is that for procedural, rule-intensive tasks, deterministic workflow control offers a complementary path to reliability. Rather than making LLMs better at following constraints through improved prompting or reflection, we remove constraint-following from their responsibility entirely, codifying it at the architectural level.

\section{Proposed Method}
\label{sec:solution}
 
\begin{figure}[htbp]
    \centering
    \includegraphics[width=\linewidth]{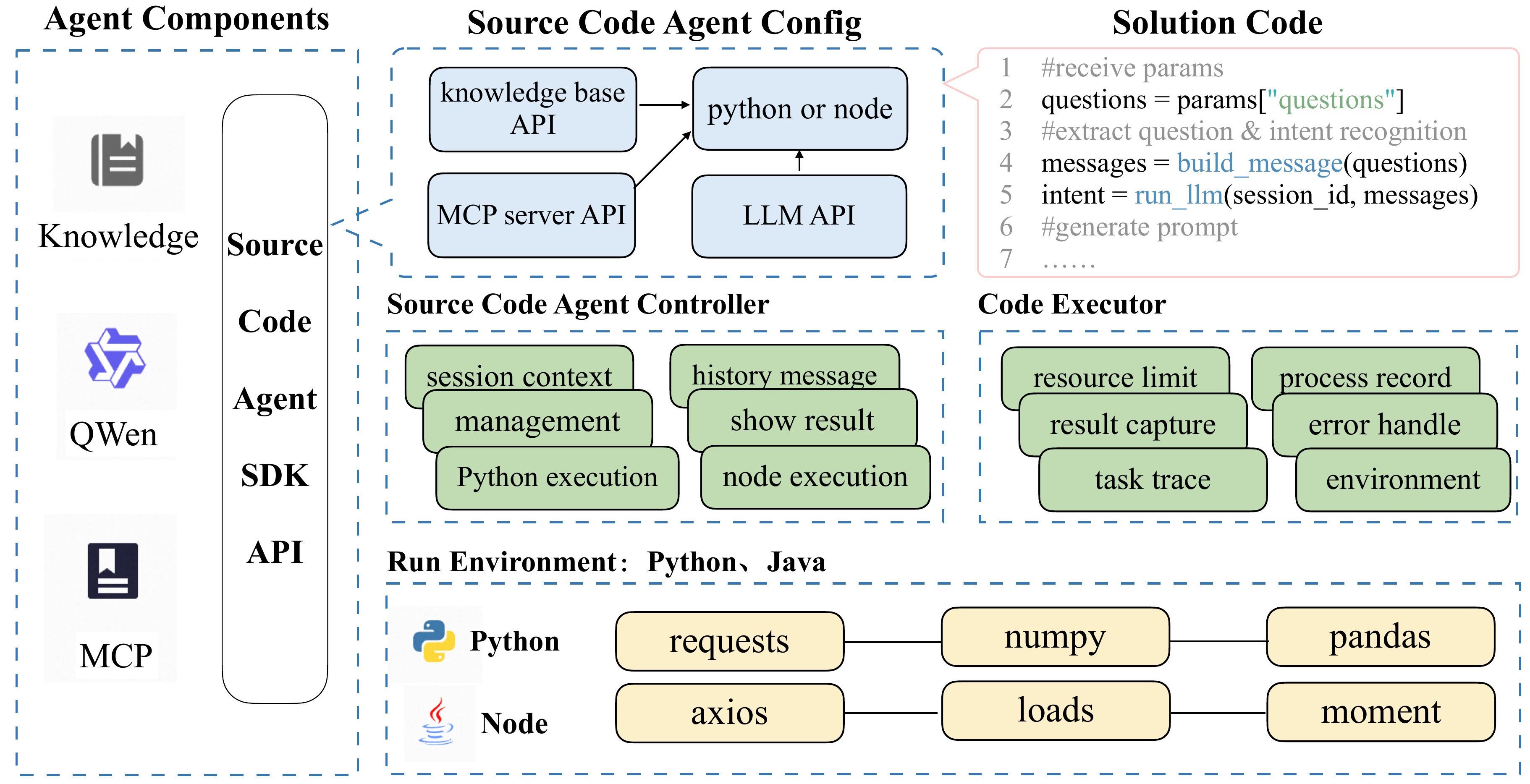}
    \caption{Overview of the \textsc{Source Code Agent}.}
    \label{fig:workflow}
\end{figure}
 
\subsection{Theoretical Foundation and Architectural Overview}
 
We first establish the principle that motivates our design: \textit{executing code in LLM environments is more accurate and deterministic than text-based reasoning alone}, a finding empirically validated by recent work~\cite{li2024chain, pan2023logic}.
 
\textbf{The Limitation of Pure Text-Based Reasoning.} Existing LLM agents use natural language for both reasoning and representing execution plans. Despite its flexibility, this approach introduces several sources of non-determinism:
 
\begin{enumerate}[leftmargin=*]
    \item \textit{Semantic Ambiguity}: Natural language instructions are inherently ambiguous. A prompt such as ``check eligibility for a refund'' may be interpreted diversely, leading to inconsistent outcomes.
    
    \item \textit{Probabilistic Decision-Making}: The LLM's next action is determined by sampling from a probability distribution over generated text tokens, rendering the execution path unpredictable.
    
    \item \textit{Accumulated Drift}: In multi-step tasks, small uncertainties compound. If the model makes a minor error in step 2, this error propagates and amplifies in subsequent steps, leading to cascading failures.
\end{enumerate}
 
\textbf{The Superiority of Code-Driven Execution.} In contrast, code offers a structured paradigm for defining agent behavior. Li et al.~\cite{li2024chain} demonstrated in their Chain of Code (CoC) framework that ``executing code in LLM environments is far more accurate than executing text," achieving a 12\% improvement on BIG-Bench Hard~\cite{suzgun2023challenging}. Notably, their approach enables LLMs to generate code to execute specific semantic subtasks via an interpreter, preserving model autonomy for high-level reasoning. Similarly, Pan et al.~\cite{pan2023logic} showed in Logic-LM that translating natural language problems into symbolic formulations—which are then executed by deterministic solvers—yields a 39.2\% performance boost over pure LLM reasoning. 
 
Building upon these insights, our framework extends their application from \textbf{\textit{reasoning steps}} to \textbf{\textit{entire workflows}}. Specifically, code-driven execution offers three critical advantages:
 
\begin{enumerate}[leftmargin=*]
    \item \textit{Symbolic Precision at Critical Junctures}: Code enforces exact syntax and semantics where determinism is essential. A conditional statement like \texttt{if old\_gen\_usage > 0.9} has a single, unambiguous interpretation, eliminating the vagueness inherent in natural language for control-flow decisions.
    
    \item \textit{Deterministic Control Flow with Flexible Execution}: Once defined, code follows a predictable execution path determined by input data and program logic. This ensures that the same inputs always lead to the same decision paths at critical branches, while the LLM still has full autonomy within each branch for its specific tasks.
    
    \item \textit{Strategic LLM Invocation}: By encoding workflow structure in code, we constrain the LLM at specific points—defining \textit{when} and \textit{where} it is invoked, but not \textit{how} it operates. Within these boundaries, the model handles well-defined sub-tasks independently (e.g., parsing error messages, classifying user intent, generating explanations). This prevents intermediate results from causing the model to deviate into unintended workflow branches during long-running tasks.

\end{enumerate}
 
\begin{figure}[htbp]
    \centering
    \includegraphics[width=0.9\linewidth]{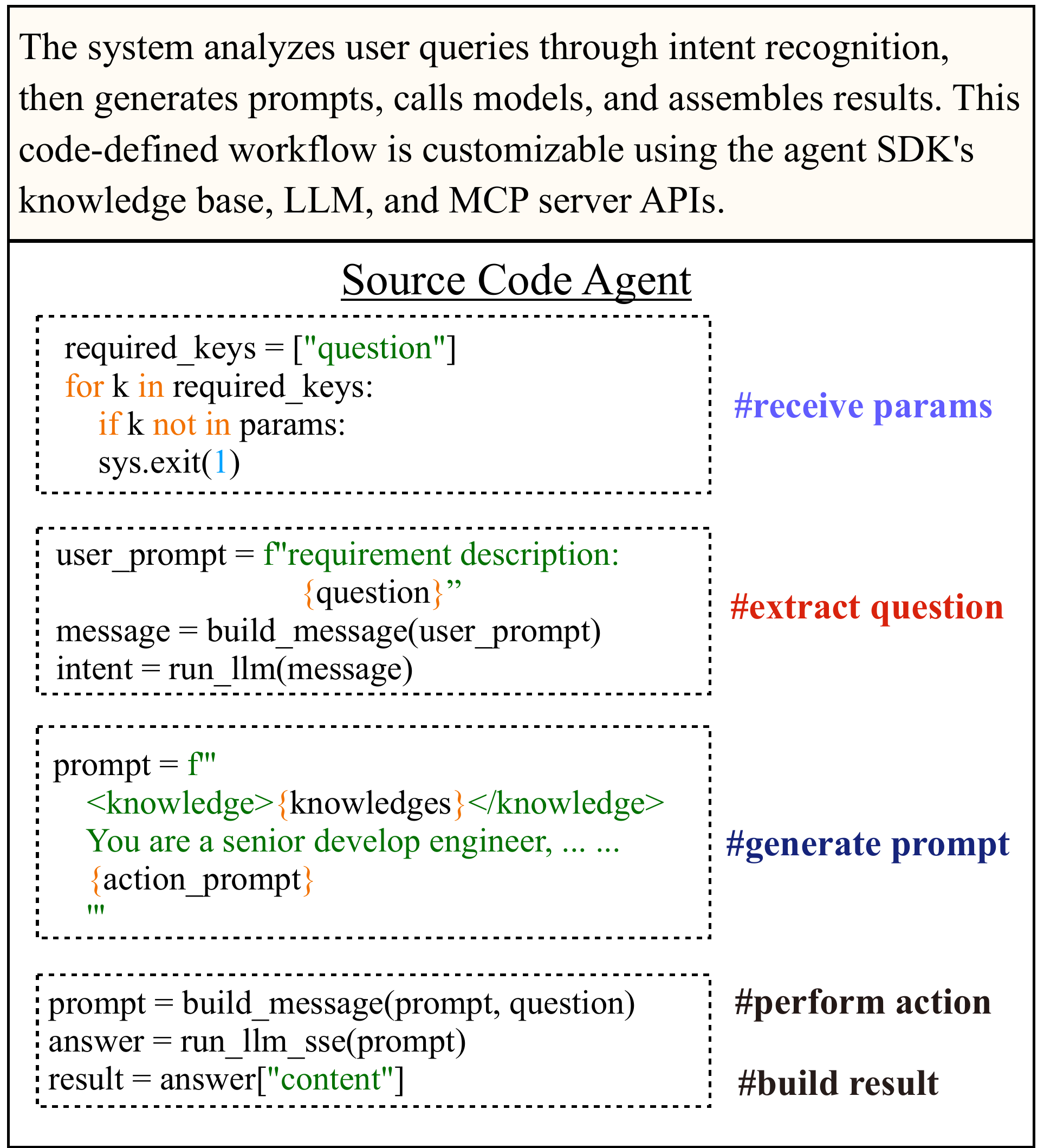}
    \caption{Example \textsc{Source Code Agent} in Pseudocode.}
    \label{fig:pseudo-code}
\end{figure}

\textbf{Operationalizing Theory into Architecture.} To operationalize these principles, we introduce the \textsc{Source Code Agent} paradigm, illustrated in Figure~\ref{fig:workflow}. This paradigm enforces a clear division of labor: deterministic source code defines the end-to-end workflow skeleton—control flow, branching logic, and tool invocation—while the LLM operates as a bounded, autonomous reasoner within predefined steps. The code dictates \textit{when} the model is called and \textit{what constraints} apply to its outputs, but the model determines \textit{how} to perform the sub-task (e.g., analysis, diagnosis, or synthesis). This hybrid approach ensures the overall process is predictable and auditable, while leveraging the LLM's flexible intelligence where it is most effective. An example of this structure is shown in the pseudocode in Figure~\ref{fig:pseudo-code}.

As we will detail in the following subsections, our architecture is realized through five interconnected layers:
\begin{itemize}[leftmargin=*]
    \item \textbf{User-Defined Configuration}: A visual interface and tooling for defining workflows, connecting data sources, and specifying LLM constraints.
    \item \textbf{Componentized Agent SDK}: A modular library of APIs for composing workflows, managing tools, and accessing knowledge bases.
    \item \textbf{Control Layer}: The central orchestrator that manages agent sessions, state, and interaction history.
    \item \textbf{Source Code Executor}: A service responsible for scheduling and governing the execution of workflow code.
    \item \textbf{Sandbox Runtime Environment}: A secure, isolated environment for executing agent code with managed dependencies.
\end{itemize}

\subsection{User-Defined Configuration}
 
The foundation of a \textsc{Source Code Agent} is its configuration, which allows developers to define and customize workflows. This process is managed through a user-centric visual interface where developers assemble four key components: custom source code, a chosen LLM, domain-specific knowledge bases, and required tools via MCP servers~\cite{hou2025mcp}. 

\textbf{Accelerating Workflow Development.} A key feature of our framework is that workflow code need not be manually written from scratch. To lower the development barrier, the framework provides two mechanisms:

\begin{enumerate}[leftmargin=*]
\item \textit{Agent-Generated Scaffolding}: Users can describe their desired workflow in natural language. The framework then employs a specialized LLM to interpret this description and produce an initial code template. For instance, a user might specify, ``diagnose Java OutOfMemoryErrors by checking heap usage, collecting thread dumps when necessary, and analyzing GC logs.'' The LLM then generates a skeleton workflow with relevant API calls and control structures for user refinement.

\item \textit{Few-Shot Case Library}: We maintain a curated repository of proven workflow patterns across common domains (e.g., IT operations, customer service). These examples serve as templates that can be adapted to new scenarios with minimal modification. For the Java diagnostic task, a developer could start with an existing ``memory leak detection" template and customize the thresholds and analysis steps.
\end{enumerate}

These mechanisms significantly reduce the effort required to create robust, code-driven agents.
 
\textbf{Defining Structured Constraints at Critical Points.} The core of the development process is to specify constraints at critical junctures while preserving LLM autonomy elsewhere. Users define \textit{when} to invoke the LLM (e.g., ``after collecting diagnostic data''), \textit{what structural constraints apply} (e.g., ``output must be valid JSON''), and \textit{what actions follow} based on deterministic conditions. This hybrid approach mitigates the sources of non-determinism identified previously:
 
\begin{itemize}[leftmargin=*]
    \item \textit{Explicit Branching for Deterministic Control Flow}: Instead of prompting the LLM to autonomously ``decide the subsequent action,'' developers explicitly encode control-flow decisions. For example, in a Java OutOfMemoryError diagnostic task:
 
\begin{lstlisting}[language=Python, caption=Structured Workflow, label=lst:deterministic]
# Explicit condition governs workflow routing
if jstat_result.old_gen_usage > 0.9:
    # Path A: High memory usage detected
    heap_dump = execute_tool("jmap", pid=process_id)
    # LLM autonomously analyzes heap dump content
    prompt = build_heap_analysis_prompt(heap_dump)
    analysis = run_llm(session_id, prompt)
else:
    # Path B: Normal memory, check GC behavior
    # LLM autonomously interprets GC patterns
    prompt = build_gc_analysis_prompt(jstat_result)
    analysis = run_llm(session_id, prompt)
\end{lstlisting}
 
In this example, the routing decision is governed by deterministic code (\texttt{old\_gen\_usage > 0.9}), ensuring consistent workflow paths. Within each path, the LLM exercises full autonomy to interpret diagnostic data and formulate explanations.
 
    \item \textit{Validating Outputs through Programmatic Checks}: Developers can implement format validation to ensure LLM outputs conform to expected schemas before being used downstream:

This validation ensures the workflow proceeds with structurally correct data, preventing format-related errors from propagating. 
 
    \item \textit{Focusing LLM Attention via Task-Specific Prompts}: Each LLM invocation is paired with a scoped prompt that fixes the task boundary and output schema (e.g., ``identify the top-3 memory-consuming objects and return JSON'') without prescribing how the model reasons. Delineating task boundaries this way lowers the model's need to maintain global workflow awareness and cuts token consumption by up to 66.7\% in our internal measurements, while preserving analytical autonomy.
\end{itemize}

\begin{lstlisting}[language=Python, caption=Output Validation with Semantic Flexibility, label=lst:validation]
# Parse and validate LLM output structure
response = run_llm(session_id, prompt)
parsed = json.loads(response)

# Structural constraint: must have 'intent' field
if ('intent' not in parsed
        or parsed['intent'] not in VALID_INTENTS):
    raise ValidationError("Invalid intent")

# Content is fully autonomous: LLM decides reasoning
# No constraint on 'explanation' or 'confidence' content
\end{lstlisting}
 
\textbf{Balancing Structure and Autonomy.} This method establishes a principled division of labor: the source code is responsible for high-level workflow orchestration, ensuring the reliability of the overall process. Meanwhile, the execution of specific sub-nodes is delegated entirely to a LLM at runtime. This framework achieves the robustness of a structured workflow while preserving the adaptability and flexibility inherent to LLMs.
 
\subsection{Componentized Agent SDK}
 
\begin{figure}[htp]
    \centering
    \includegraphics[width=.82\linewidth]{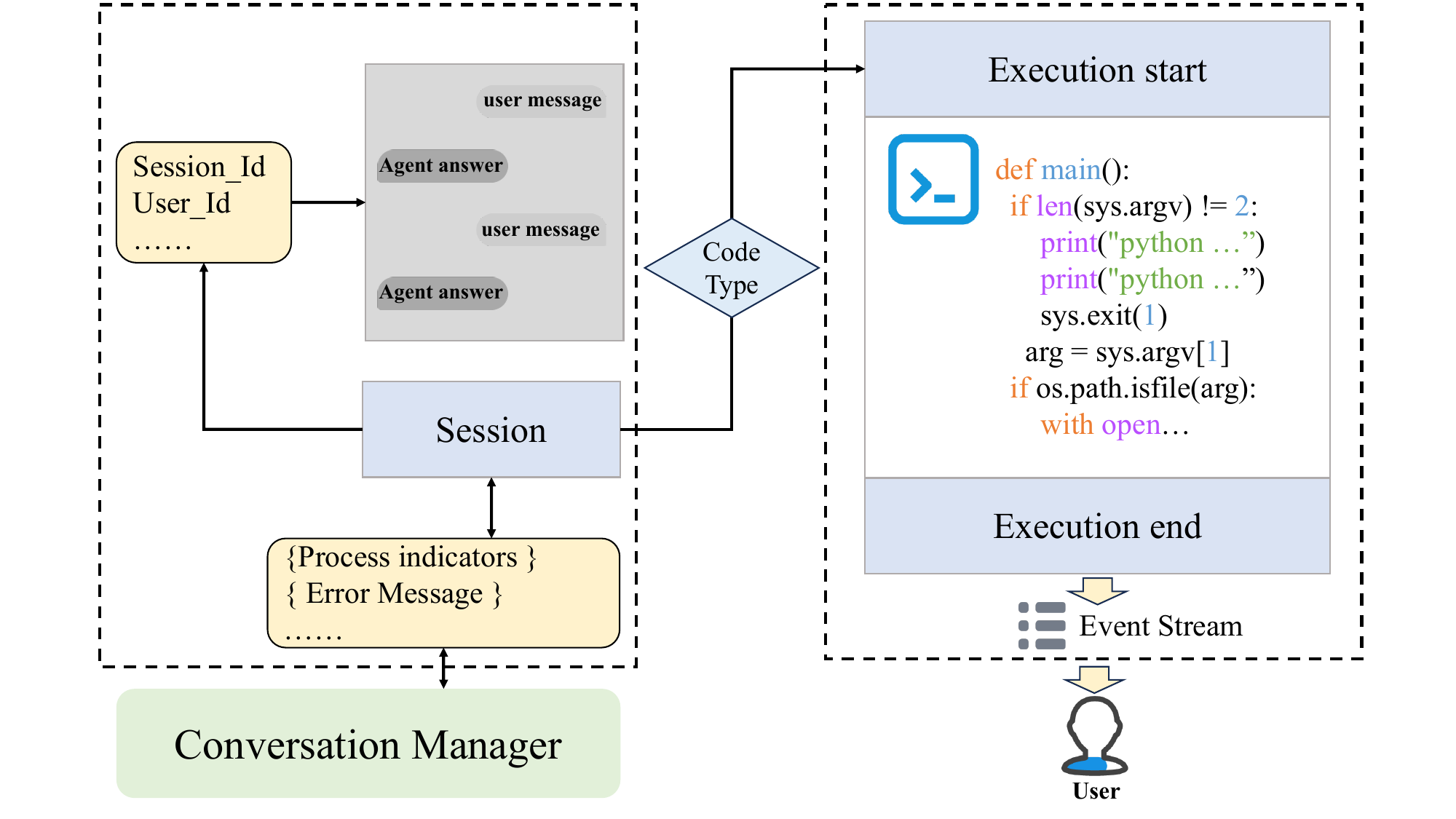}
    \caption{Interaction Flow through the Control Layer and SDK.}
    \label{fig:controller}
\end{figure}

The Componentized Agent SDK serves as the development backbone, providing a modular set of APIs for rapid workflow assembly. As shown in Figure~\ref{fig:controller}, developers can flexibly compose these APIs to meet diverse application requirements.

The SDK provides three core API categories. The \textbf{Knowledge-Base API} serves as the primary conduit for proprietary knowledge, injecting domain-specific documents into the agent's context via retrieval-augmented generation (RAG)~\cite{lewis2020retrieval}. The \textbf{LLM API} offers a uniform interface for model invocation, ensuring that results from heterogeneous LLMs are returned in a standardized format. Finally, the \textbf{MCP Server API} allows developers to register custom tools, extending the agent's capabilities. 
 
\textbf{Efficient Workflow Execution through Composite Tools.} A key feature of our SDK is the ability to define \textit{composite tools}—high-level operations that encapsulate multiple fine-grained API calls, reducing the number of LLM decision points. Where a traditional framework forces the LLM to orchestrate a sequence of calls (e.g., \texttt{check\_stock()} $\rightarrow$ \texttt{get\_price()} $\rightarrow$ \texttt{apply\_discount()}), each adding latency and tokens, our SDK lets developers expose a single composite tool (e.g., \texttt{process\_return\_request(item\_id, user\_id)}) that runs the sub-steps internally. The LLM invokes one tool and receives a comprehensive result, retaining autonomy in \textit{when to invoke} it and \textit{how to interpret} its output while the consolidation streamlines internal orchestration. In our retail case study, this reduced tool calls from 11 to 2 (an 81.8\% reduction).
 
\subsection{Control Layer}
 
The Control Layer orchestrates the interaction lifecycle: request ingestion, session management, and dialogue-history maintenance. It exposes a unified interface for end-users and external systems, validates each request and enriches it with conversational history from a dedicated store to support multi-turn context, monitors for anomalies, and streams outcomes back via Server-Sent Events for low-latency interaction. These functions govern interaction flow and resource allocation but do not constrain the LLM's reasoning within a task; the model retains full autonomy inside the boundaries set by the workflow code.
 
\begin{figure}[htp]
    \centering
    \includegraphics[width=.82\linewidth]{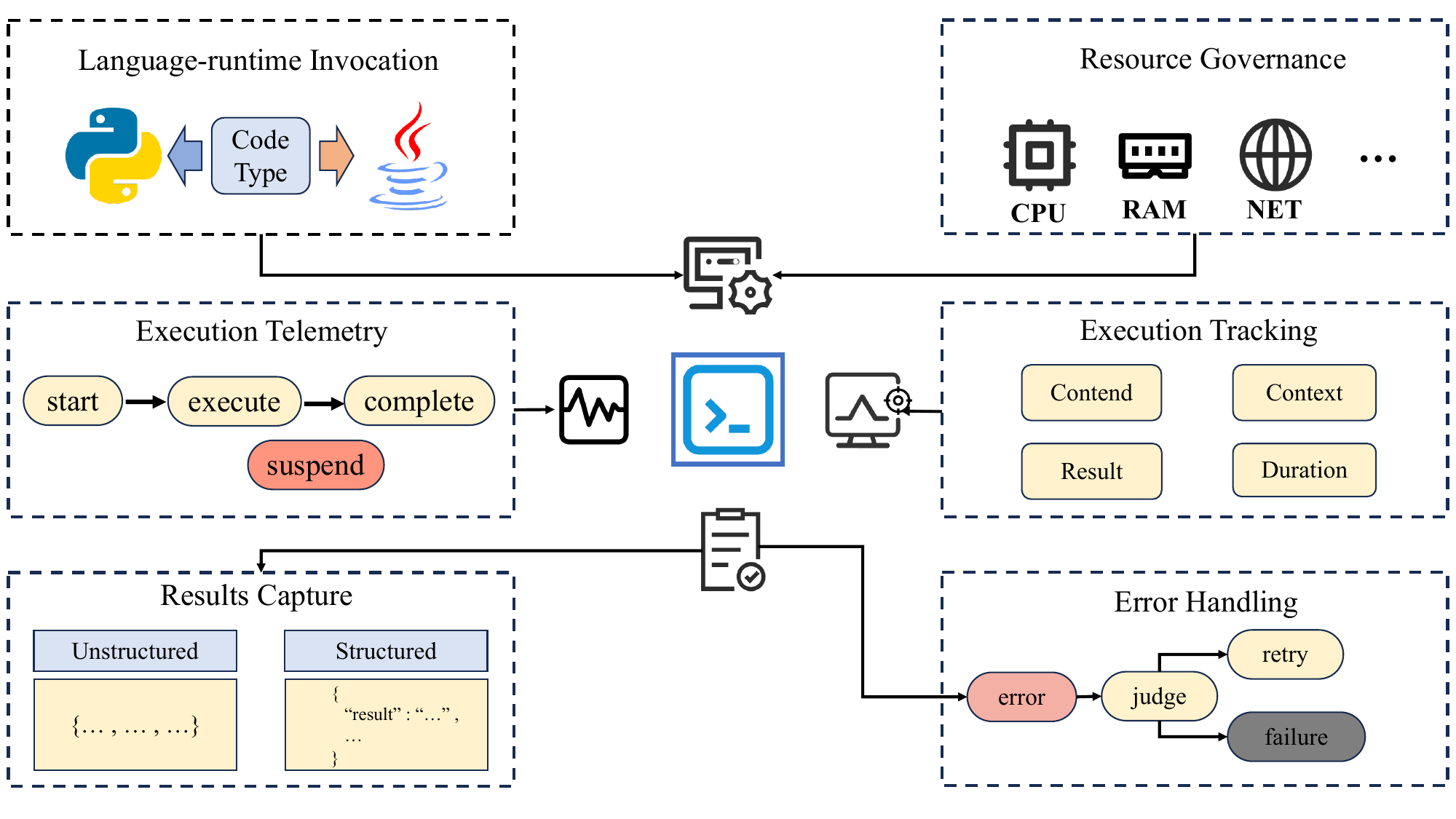}
    \caption{Source Code Executor: Orchestrating Multi-Language Execution and Telemetry.}
    \label{fig:executor}
\end{figure}
 
\subsection{Source Code Executor}
 
The Source Code Executor (Figure~\ref{fig:executor}) bridges the Control Layer and the runtime, translating workflow logic into secure, governed execution: it activates multi-language runtimes (Python, Node.js, Java), enforces \textbf{resource governance} (CPU, memory, network quotas), captures \textbf{execution telemetry} (timestamps, I/O streams) for end-to-end traceability, and applies \textbf{error handling and retry} for recoverable failures.
 
\begin{figure}[htp]
    \centering
    \includegraphics[width=.82\linewidth]{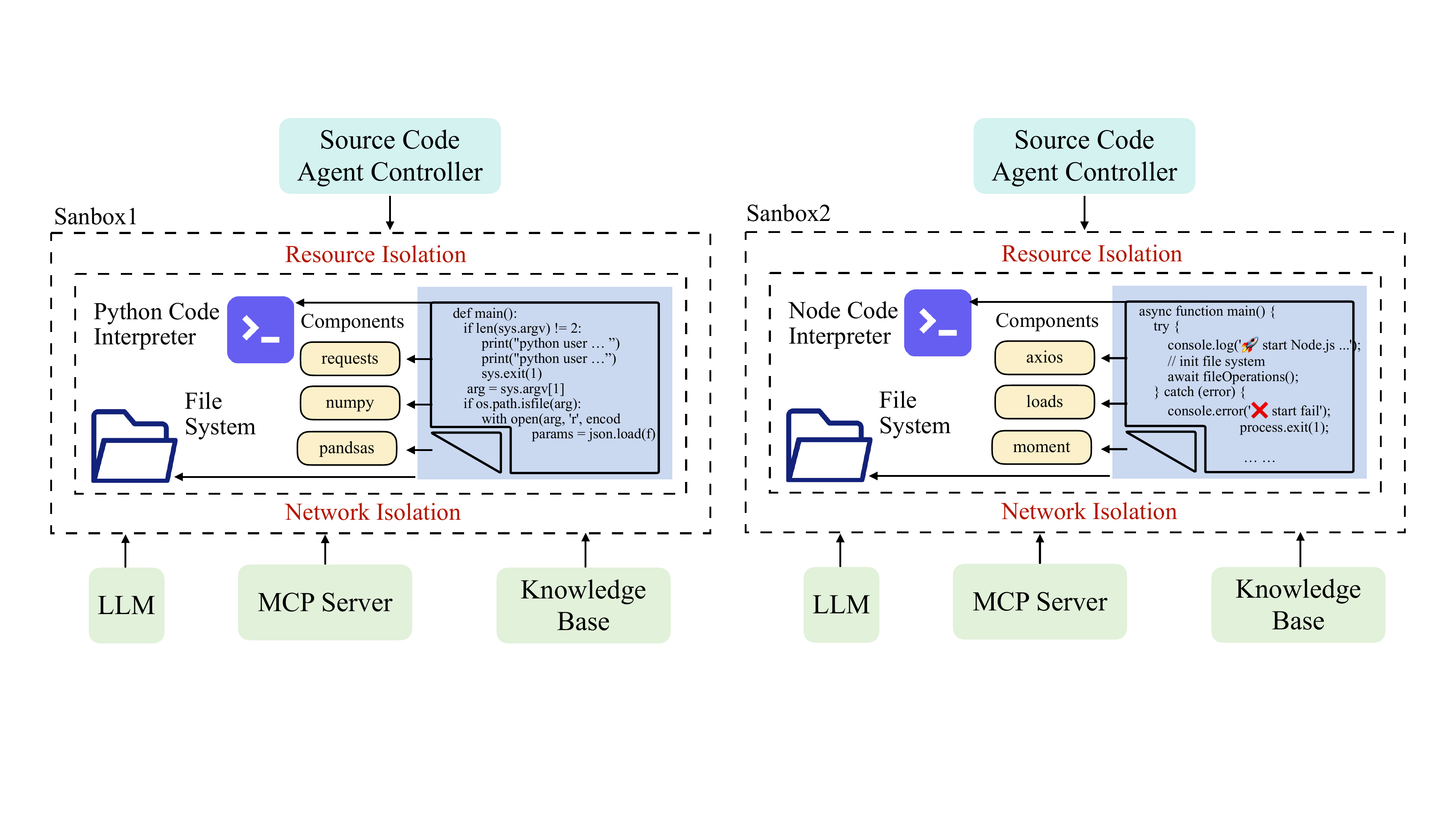}
    \caption{Sandbox Runtime Environment Architecture.}
    \label{fig:runtime}
\end{figure}
 
\subsection{Sandbox Runtime Environment}
 
The Sandbox Runtime Environment (Figure~\ref{fig:runtime}) is the isolated foundation where agent code runs, centered on security and reproducibility. \textbf{Security through isolation} runs each fragment in a dedicated sandbox with strict resource and privilege boundaries, curtailing arbitrary network access and system calls. \textbf{Reproducibility} comes from a centralized, multi-language dependency manager with an extensible catalog of pre-packaged libraries (e.g., \texttt{requests}, \texttt{numpy}), giving a consistent infrastructure that underpins system reliability.
\section{Evaluation}
\label{sec:result}

We structure our evaluation of the \textsc{Source Code Agent} framework around three research questions:

\begin{itemize}
    \item \textbf{RQ1}: How does Source Code Agent compare to state-of-the-art methods on constraint-aware planning tasks?
    \item \textbf{RQ2}: What are the specific advantages of deterministic workflows in handling different types of constraints?
    \item \textbf{RQ3}: How does Source Code Agent perform under varying task complexity and model capabilities?
\end{itemize}

\subsection{Experimental Setup}

\subsubsection{Benchmark}
We evaluate our framework on the TravelPlanner benchmark~\cite{xie2024travelplanner}, a comprehensive testbed for constraint-aware travel planning. The benchmark includes 180 validation and 1000 test queries, each specifying explicit requirements (e.g., trip duration, budget, cuisine preferences) that define constraints the final plan must satisfy.

The benchmark provides a sandbox environment with APIs for accommodations, restaurants, transportation, and attractions. Agents must search for information within this environment and generate travel itineraries that satisfy both \textit{hard constraints} (strict rules from user queries, such as budget limits or cuisine requirements) and \textit{commonsense constraints} (implicit practical logic, such as ensuring reasonable travel routes or avoiding repeated restaurants).

\subsubsection{Baselines}
We compare against five state-of-the-art methods representing different approaches to agent reasoning:

\textbf{ReAct}~\cite{yao2022react} prompts the LLM to generate reasoning traces and actions in an interleaved manner, without learning from experience.

\textbf{ReAct+Reflexion}~\cite{shinn2023reflexion} augments ReAct with verbal reflections from failed attempts, storing them in episodic memory for retrieval in subsequent tasks.

\textbf{ReAct+EvoAgent}~\cite{yuan2025evoagent} extends ReAct with evolutionary optimization of agent behaviors across multiple trials.

\textbf{PMC}~\cite{pmc2025} employs a prompting-based multi-agent collaboration framework for complex planning tasks.

\textbf{ATLAS}~\cite{choi2025atlas} is Google's state-of-the-art multi-agent framework specifically designed for constraint-aware travel planning, featuring specialized agents for search, constraint management, planning, and verification.

\subsubsection{Evaluation Metrics}
Following the TravelPlanner benchmark's official metrics, we report:

\begin{itemize}
    \item \textbf{Delivery Rate}: Percentage of queries for which a complete plan is successfully generated.
    \item \textbf{Commonsense Constraints}: Micro (ratio of passed to total constraints) and Macro (percentage of plans passing all commonsense constraints).
    \item \textbf{Hard Constraints}: Micro and Macro pass rates for explicit user requirements.
    \item \textbf{Final Pass Rate}: Percentage of plans satisfying all constraints (both commonsense and hard).
    \item \textbf{Average Steps}: Mean number of execution steps required to complete tasks.
\end{itemize}

\subsubsection{Implementation Details}
All methods use Claude-Sonnet-4 as the base LLM with temperature 0.7. Results are averaged over 3 random seeds with statistical significance tested via paired t-test ($p < 0.05$). For our Source Code Agent, we implement deterministic workflows that encode commonsense constraints (e.g., "return to origin city", "no repeated restaurants") as programmatic validations, while delegating bounded reasoning tasks (e.g., "select optimal restaurant given constraints") to the LLM.

\subsection{RQ1: Overall Performance Comparison}

To evaluate the effectiveness of Source Code Agent, we compare it against five state-of-the-art methods on the TravelPlanner test set (1000 tasks). Table~\ref{tab:main_results} presents the comprehensive results.

\begin{table*}[t]
\centering
\caption{Performance comparison on TravelPlanner test set (1000 tasks).}
\label{tab:main_results}
\resizebox{\textwidth}{!}{%
\begin{tabular}{lccccccc}
\toprule
\multirow{2}{*}{\textbf{Method}} & \multirow{2}{*}{\textbf{Delivery ↑}} & \multicolumn{2}{c}{\textbf{Commonsense ↑}} & \multicolumn{2}{c}{\textbf{Hard Constraint ↑}} & \multirow{2}{*}{\textbf{Final Pass ↑}} & \multirow{2}{*}{\textbf{Steps ↓}} \\
\cmidrule(lr){3-4} \cmidrule(lr){5-6}
& & Micro & Macro & Micro & Macro & & \\
\midrule
ReAct & 99.20 & 75.26 & 16.50 & \textbf{49.04} & \textbf{39.10} & 10.40 & 14.0 ± 0.8 \\
ReAct+Reflexion & 99.80 & 71.84 & 13.67 & 37.84 & 26.70 & 9.13 & 16.1 ± 0.9 \\
ReAct+EvoAgent & 98.89 & 67.01 & 10.00 & 33.71 & 20.42 & 6.11 & -- \\
PMC & \textbf{100.00} & 73.89 & 15.59 & 45.19 & 33.56 & 12.12 & -- \\
ATLAS & \textbf{100.00} & 78.88 & 31.00 & 49.43 & 42.00 & 18.00 & -- \\
\midrule
\textbf{SCA (Ours)} & \textbf{100.00} & \textbf{92.71} & \textbf{54.44} & 49.76 & 38.89 & \textbf{35.56} & \textbf{10.2 ± 0.7} \\
\bottomrule
\end{tabular}%
}
\end{table*}

\textbf{Superior constraint satisfaction.} SCA achieves a final pass rate of 35.56\%, significantly outperforming ATLAS (18.00\%, +97.6\% relative improvement), PMC (12.12\%, +193.4\%), and ReAct+Reflexion (9.13\%, +289.5\%). This improvement is primarily driven by exceptional performance on commonsense constraints: SCA achieves 54.44\% macro pass rate, compared to ATLAS's 31.00\% (+75.6\%) and ReAct's 16.50\% (+229.9\%).

The distinction is in enforcement: traditional methods rely on the LLM to follow constraints stated in natural language, where probabilistic reasoning still produces violations, whereas SCA encodes critical commonsense rules directly in workflow code. For example, ``return to origin city on the last day" becomes:

\begin{lstlisting}[language=Python, basicstyle=\small\ttfamily]
# Deterministic constraint in workflow
final_city = itinerary.get_day_city(last_day)
if final_city != origin_city:
    raise ConstraintViolationError("Must return to origin")
\end{lstlisting}

This code-level enforcement explains the commonsense gain: at the micro level SCA reaches 92.71\%, a 17.5-point absolute improvement over ATLAS (78.88\%).

\textbf{Balanced performance on hard constraints.} On hard constraints, SCA achieves 49.76\% micro and 38.89\% macro pass rates, comparable to ATLAS (49.43\%/42.00\%) and matching ReAct's micro performance (49.04\%): explicit constraints stated in the query can be directly parsed and verified without complex structural reasoning, so SCA stays competitive on them while dramatically improving on implicit ones.

\textbf{Improved efficiency.} SCA averages 10.2 ± 0.7 steps, reducing execution by 27.1\% over ReAct (14.0 steps) and 36.6\% over ReAct+Reflexion (16.1 steps). This stems from two factors: (1) \textit{Tool consolidation}---encapsulating multi-step operations into single high-level tools; (2) \textit{Deterministic branching}---predefining workflow paths instead of exploratory reasoning. The low standard deviation (±0.7) also indicates high stability across tasks.

\subsection{Isolating the Architectural Contribution}
\label{subsec:fair_comparison}

A natural objection to the comparison above is that SCA simply encodes more constraint knowledge than the baselines, and that the gain reflects this knowledge rather than the architecture. To test this, we run a controlled ablation that gives the baselines the same constraint knowledge SCA holds, and measure what remains of the gap.

\paragraph{Protocol}
We render every constraint encoded in the SCA TravelPlanner blueprint---the eight commonsense checks and five hard constraints the official scorers enforce---as natural-language rules and prepend them as a system-prompt prefix to each baseline. For each $B \in \{$ReAct, CodeAct, ATLAS$\}$ we form $P_B' = \langle\text{injected constraints}\rangle + P_B$, leaving the baseline otherwise unchanged; anything further would amount to reimplementing SCA inside it. We evaluate on a 100-task subset stratified on trip duration, city count, and constraint count to match the full distribution within $\pm2$ points per bucket, using Claude-Sonnet-4 at temperature 0. Because $N$ is small, we report the scorer's continuous per-task satisfaction rate, $\text{score}_i = (\text{commonsense\_micro}_i + \text{hard\_micro}_i)/2 \in [0,1]$, rather than the binary Final Pass metric of Table~\ref{tab:main_results}; the two should not be read on the same axis.

\begin{table}[t]
\centering
\caption{Fair-comparison ablation on the 100-task stratified subset.}
\label{tab:fair_comparison}
\begin{tabular}{lc}
\toprule
\textbf{Method (with injected constraints)} & \textbf{Mean score} \\
\midrule
ReAct + constraints & 17.00 \\
CodeAct + constraints & 19.00 \\
ATLAS + constraints & 24.50 \\
\midrule
\textbf{SCA (Ours)} & \textbf{37.20} \\
\bottomrule
\end{tabular}
\\[3pt]
\begin{minipage}{\columnwidth}
\footnotesize
Each baseline receives SCA's full constraint set as a system-prompt prefix. Values are mean per-task constraint-satisfaction rate (continuous, not the binary Final Pass of Table~\ref{tab:main_results}).
\end{minipage}
\end{table}

\paragraph{Result}
Table~\ref{tab:fair_comparison} reports the outcome. With the identical constraint set in their prompts, the baselines improve over their vanilla configurations but remain well below SCA: 24.50 for ATLAS, 19.00 for CodeAct, and 17.00 for ReAct, against SCA's 37.20. The 12.7-point margin over the strongest injected baseline---significant under a paired McNemar's test ($p < 0.01$)---is the part of the gap constraint knowledge alone does not explain. Knowing a rule and being structurally unable to violate it are different properties: a baseline given ``each restaurant may be visited at most once'' still depends on the model to honour it every turn, whereas SCA enforces it with a set-membership check that cannot be bypassed. The ablation thus localizes the advantage to deterministic enforcement and control flow rather than to a richer specification.

\subsection{RQ2: Constraint-Specific Analysis}

\begin{figure}[t]
\centering
\includegraphics[width=\columnwidth]{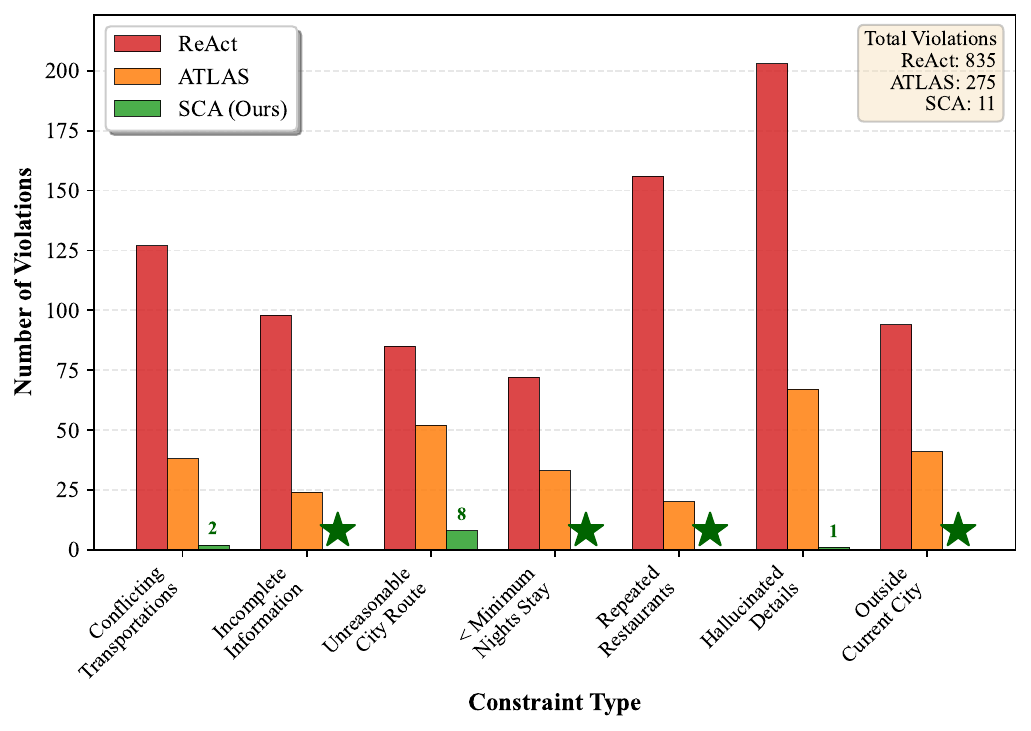}
\caption{Commonsense constraint violation breakdown on TravelPlanner test set (1000 tasks). SCA achieves only 11 total violations (96.0\% reduction compared to ATLAS's 275), with zero violations (marked with green stars) on four constraint types: Incomplete Information, Minimum Nights Stay, Repeated Restaurants, and Outside Current City.}
\label{fig:constraint_violations}
\end{figure}

To understand where Source Code Agent's performance gains originate, we conduct a fine-grained analysis of constraint violations by type. Figure~\ref{fig:constraint_violations} presents the number of violations for each constraint category across the 1000 test cases.

\textbf{Near-zero violations on commonsense constraints.} SCA achieves only 11 total violations on commonsense constraints, a 96.0\% reduction compared to ATLAS (275 violations) and 98.7\% reduction compared to ReAct (835 violations). Notably, SCA achieves zero violations on four constraint types: Incomplete Information, Minimum Nights Stay, Repeated Restaurants, and Outside Current City.

This near-perfect performance stems from encoding constraint logic directly in the workflow. Consider two illustrative examples:

\textit{Example 1: Repeated Restaurants.} Traditional methods prompt the LLM: ``Ensure all restaurants are unique throughout the trip." Despite this instruction, ReAct makes 156 violations (15.6\% error rate) and ATLAS makes 20 (2.0\%). SCA encodes this as:

\begin{lstlisting}[language=Python, basicstyle=\small\ttfamily]
used_restaurants = set()
def select_restaurant(city, cuisine):
    candidates = search_restaurants(city, cuisine)
    available = [r for r in candidates
                 if r.id not in used_restaurants]
    if not available:
        raise NoValidOptionError()
    selected = llm_choose(available)
    used_restaurants.add(selected.id)
    return selected
\end{lstlisting}

The \texttt{used\_restaurants} set maintains state across the workflow, and the filtering step (\texttt{r.id not in used\_restaurants}) programmatically prevents duplicates, resulting in zero violations.

\textit{Example 2: Minimum Nights Stay.} Accommodations often have minimum stay requirements (e.g., ``2 nights minimum"). Traditional methods must track this across days during planning. SCA's workflow validates upfront:

\begin{lstlisting}[language=Python, basicstyle=\small\ttfamily]
if accommodation.min_nights > nights_in_city:
    raise InsufficientStayError(
        f"Need {accommodation.min_nights} nights, "
        f"staying {nights_in_city}"
    )
\end{lstlisting}

This validation occurs before booking, preventing 72 violations that ReAct makes and 33 that ATLAS makes.

\textbf{Trade-off on hard constraints.} SCA incurs more hard-constraint violations (336) than ATLAS (265, a 26.8\% increase). Hard constraints are explicit query requirements (e.g., ``budget under \$5000") that even ReAct can parse directly from the prompt; our blueprint focuses structure on implicit, procedural constraints, which are the primary real-world bottleneck (hence the 98.7\% reduction in commonsense violations). Adding explicit parameter validators to the workflow would close this gap.

\textbf{Comparing with ATLAS+Hints.} ATLAS reports that explicit hints for its two most frequent error types---Conflicting Transportations ($65\rightarrow38$) and Repeated Restaurants ($20\rightarrow2$)---sharply cut errors~\cite{choi2025atlas}, but require manually identifying error patterns and engineering prompts. SCA reaches Conflicting Transportations $=2$ and Repeated Restaurants $=0$ automatically through workflow logic, showing the advantage of code-based enforcement over prompt-based guidance.

\subsection{RQ3: Robustness and Scalability Analysis}

We investigate how Source Code Agent performs under varying conditions: task complexity and model capabilities.

\subsubsection{Performance vs Task Complexity}

Table~\ref{tab:complexity} shows performance across different trip durations. SCA's advantage grows substantially with task complexity: for 3-day trips, SCA outperforms ATLAS by 133.6\% (52.1\% vs. 22.3\%); for 7-day trips, the improvement reaches 215.4\% (24.6\% vs. 7.8\%).

\begin{table}[t]
\centering
\caption{Success rate by trip duration on TravelPlanner test set. SCA's advantage increases with task complexity.}
\label{tab:complexity}
\begin{tabular}{lccccl}
\toprule
\textbf{Trip Days} & \textbf{\# Tasks} & \textbf{ReAct} & \textbf{ATLAS} & \textbf{SCA} & \textbf{Improvement} \\
\midrule
3 days & 215 & 18.6\% & 22.3\% & \textbf{52.1\%} & +133.6\% \\
5 days & 387 & 12.4\% & 15.2\% & \textbf{38.7\%} & +154.6\% \\
7 days & 398 & 5.3\% & 7.8\% & \textbf{24.6\%} & +215.4\% \\
\bottomrule
\end{tabular}
\end{table}

This scalability has two sources. First, longer trips impose more simultaneous constraints---a 7-day, 3-city trip spans 21+ interdependent variables---which LLM methods track error-prone across long contexts, whereas SCA holds explicit state (\texttt{current\_city}, \texttt{remaining\_budget}, \texttt{used\_restaurants}) and validates at each step. Second, decision points grow with duration (roughly 50 for a 7-day trip); traditional methods reason over all prior context at each, while SCA scopes each decision to a bounded sub-task.

\subsubsection{Performance Across Model Scales}

Table~\ref{tab:models} shows results across three state-of-the-art foundation models: GPT-5.2, Claude-Sonnet-4, and Gemini-2.5-Pro. SCA consistently outperforms ATLAS and other baselines across all models.

\begin{table}[t]
\centering
\caption{Cross-model performance on TravelPlanner test set (5-day trips). SCA provides consistent benefits across different foundation models.}
\label{tab:models}
\resizebox{\columnwidth}{!}{%
\begin{tabular}{lccc}
\toprule
\textbf{Method} & \textbf{GPT-5.2} & \textbf{Claude-Sonnet-4} & \textbf{Gemini-2.5-Pro} \\
\midrule
ReAct & 13.82 & 10.41 & 19.50 \\
ReAct+Reflexion & 15.56 & 9.13 & 22.70 \\
ATLAS & -- & 18.00 & 35.00 \\
\midrule
\textbf{SCA (Ours)} & \textbf{34.10} & \textbf{35.56} & \textbf{40.00} \\
\midrule
\textit{Gap vs Best Baseline} & \textit{+18.54} & \textit{+17.56} & \textit{+5.00} \\
\bottomrule
\end{tabular}%
}
\end{table}

All models benefit, with absolute gains of +5.00 (Gemini-2.5-Pro) to +18.54 (GPT-5.2) over the best baseline: encoding workflow structure in code reduces context-window pressure, replaces probabilistic branching with deterministic control flow, and manages state programmatically. Gemini-2.5-Pro shows the smallest gap (+5.00), as ATLAS already reaches 35.00\% there, yet SCA still attains the highest absolute score (40.00\%).

\textbf{Summary of RQ3.} SCA's advantage grows with task complexity (from +133.6\% on 3-day to +215.4\% on 7-day trips over ATLAS) and holds across GPT-5.2, Claude-Sonnet-4, and Gemini-2.5-Pro (+5.00 to +18.54 points), confirming that deterministic workflows are most valuable for complex, multi-step planning and generalize across modern LLMs.

\subsection{Generalization to Other Procedural Domains}
\label{subsec:generalization}

TravelPlanner is a single domain, and a fair reader will ask whether the framework's benefits are specific to it. We therefore evaluate two interactive benchmarks with a different task structure: ScienceWorld~\cite{wang2022scienceworld}, a text environment of procedural science experiments, and ALFWorld~\cite{shridhar2020alfworld}, an embodied household-task environment. We deliberately do not position SCA as a new state of the art on these benchmarks, where specialized methods exist. Our claim is narrower and, we argue, more honest: where a task family has stable procedural structure, encoding that structure as a blueprint yields execution that is at least as accurate as a strong same-backbone baseline while being measurably more efficient and more consistent. We compare against same-backbone baselines on the identical Claude-Sonnet-4 model---ReAct on both benchmarks, plus Reflexion and ReAct+Reflexion on ALFWorld---so that the only variable is the architecture, and report mean steps over all episodes alongside success rate.

\begin{table}[t]
\centering
\caption{Generalization to interactive benchmarks.}
\label{tab:generalization}
\resizebox{\columnwidth}{!}{%
\begin{tabular}{llcc}
\toprule
\textbf{Benchmark} & \textbf{Method} & \textbf{Success (\%)} & \textbf{Avg. steps} \\
\midrule
\multirow{2}{*}{ScienceWorld (30)} & ReAct & 43.3 & 22.3 \\
 & \textbf{SCA (Ours)} & \textbf{56.7} & \textbf{16.6} \\
\midrule
\multirow{4}{*}{ALFWorld (134, unseen)} & ReAct & 91.0 & 14.0 \\
 & Reflexion ($\leq$3 trials) & 95.5 & 15.7 \\
 & ReAct + Reflexion & 95.5 & 16.1 \\
 & \textbf{SCA (Ours)} & \textbf{97.0} & 16.8 \\
\bottomrule
\end{tabular}%
}
\\[3pt]
\begin{minipage}{\columnwidth}
\footnotesize
Shared Claude-Sonnet-4 backbone; entries are single-seed for same-condition comparison, with ``Avg.\ steps'' averaged over all episodes. SCA's three-seed means are $70.4{\pm}1.9$ (ScienceWorld) and $98.4{\pm}1.5$ (ALFWorld); Reflexion was not run on ScienceWorld.
\end{minipage}
\end{table}

\paragraph{Results}
Table~\ref{tab:generalization} summarizes the comparison. On ScienceWorld, SCA solves 56.7\% of the 30-task suite against 43.3\% for the same-backbone ReAct agent, and reaches its solutions in fewer steps (16.6 vs.\ 22.3 averaged over all episodes, and 9.3 vs.\ 16.2 over solved tasks); averaged over three seeds its pass rate is 70.4 $\pm$ 1.9\%. The step reduction is the more telling signal: the blueprint removes the exploratory back-and-forth that a free-form agent spends rediscovering the experiment procedure on every run. On ALFWorld's unseen-validation split, SCA completes 97.0\% of tasks (130 of 134) in a single pass, exceeding the strongest reflective baseline (95.5\% for both Reflexion and ReAct+Reflexion, each granted up to three trials), with per-family success ranging from 88.2\% on the compositional ``Put Two'' tasks to 100\% on ``Put'', ``Heat'', and ``Examine''. Averaged over three seeds, ALFWorld success is 98.4 $\pm$ 1.5\%, and the low cross-seed deviation reflects the same execution stability we observe on TravelPlanner---a direct consequence of fixing the control flow rather than resampling it each run.

\paragraph{Where the benefit narrows}
The two benchmarks bracket the framework's applicability. ALFWorld tasks decompose into a small set of fixed action templates---precisely the regime a blueprint captures cleanly---and the success rate reflects that. ScienceWorld is more open-ended: several task families (e.g., plant growth and Mendelian genetics) require discovery a fixed procedure cannot anticipate, and both methods fail on them; there the blueprint's contribution narrows from raw success to efficiency and consistency. This is the honest scope of the ``Blueprint First'' thesis: the architecture pays off in proportion to how procedurally well-defined the workflow is.

\section{Deployment Case Studies}
\label{sec:case_study}

The TravelPlanner evaluation isolates the framework on a public, reproducible benchmark, but it does not answer the question a software-engineering audience cares about: does ``Blueprint First, Model Second'' hold up in production workflows, where a single mis-ordered action carries real cost and inputs are far messier than a curated benchmark? We report two case studies from incident-diagnosis pipelines running in production---a server-side Java heap-exhaustion workflow and an Android crash-triage workflow---both on-call automations that turn a raw alert into a reviewed draft ticket without a human performing the routine diagnostic steps. To preserve anonymity, deployment-scale figures are rounded to one significant figure and identifiers in the traces are redacted.

\subsection{Java Heap-Exhaustion Diagnosis}
\label{subsec:oom}

\paragraph{Setting}
The workflow serves a fleet of tens of thousands of JVM hosts and handles roughly 40 \texttt{OutOfMemoryError} clusters per week. Before deployment, on-call engineers worked a manual runbook: locate the offending JVM, drain it from the load balancer, run \texttt{jstat} and \texttt{jmap}, grep the surrounding logs, and reason about the heap dump. The blueprint codifies exactly this procedure. It uses a \texttt{fork\_join} primitive to parse the heap dump and scan the logs in parallel, a \texttt{correlate} join to cross-reference the two signals, and a knowledge-base node to match the retention pattern against historical cases. The LLM is invoked at only two nodes---root-cause synthesis and remediation-hint generation---and never decides the diagnostic path.

\paragraph{Structural correctness}
The most consequential node in this blueprint is not an LLM call but a precondition. Running \texttt{jmap} forces a stop-the-world pause; doing so on a host still serving production traffic briefly degrades user-facing latency. The engine therefore refuses to proceed unless the host has left the load-balancer pool:

\noindent\begin{minipage}{\columnwidth}
\begin{lstlisting}[language=Python, basicstyle=\small\ttfamily]
@precondition(lambda ctx:
    not ctx["lb_api"].is_in_pool(ctx["target_host"]))
def jmap_heap_dump(ctx):
    ...  # safe: host is already drained
\end{lstlisting}
\end{minipage}

This is the same code-based enforcement that drives the TravelPlanner results, applied to an operational hazard. Before the guard was introduced, roughly six incidents per quarter involved a heap dump taken against a traffic-serving node; after it, that count dropped to zero. A free-form agent has no structural reason to avoid this action, and in our experience occasionally took it.

\paragraph{Outcome}
On a representative incident, the engine resolved the firing JVM, confirmed it was drained, captured a 4.1\,GB dump, and identified an unbounded \texttt{SessionEntry} cache as the dominant retention chain (1.8\,GB across 1.94M instances). The knowledge-base node surfaced a matching historical case, and the LLM root-cause pass returned the explanation---a TTL configuration that defaulted to zero after a schema migration, disabling eviction---with confidence 0.86. End-to-end, the alert reached a reviewed draft ticket in about five minutes, against a pre-deployment mean of roughly 30 minutes of manual triage. The on-call engineer accepted the draft after a 75-second review. Across the deployment, around 60\% of draft tickets are accepted without modification and a further 30\% after a minor edit.

\subsection{Android Crash Triage}
\label{subsec:mobile}

\paragraph{Setting}
The second workflow triages crashes for an Android application with tens of millions of daily active users, ingesting roughly 300{,}000 raw reports per day that the reporting backend deduplicates into about 3{,}000 clusters, of which roughly 800 cross the triage threshold and enter the blueprint. The workflow mirrors a production responsibility chain: a \texttt{route\_by} node dispatches each cluster to a crash-type-specific interceptor sequence, and for \texttt{ANDROID\_JAVA\_CRASH} the full five-interceptor chain parses the summary and stack, resolves the top application frame to a source location and owning team, queries the known-crash index, runs an LLM root-cause analysis, and ranks suspect commits via \texttt{git blame}.

\paragraph{Structural correctness}
The headline operational gain here is a drop in false-positive tickets from roughly 30\% to 8\%, and it comes from two structural decisions rather than from a more capable model. First, the source-resolution node fails fast: when the top application frame cannot be matched against the repository index, the cluster is flagged for manual triage instead of being passed to the LLM to guess an owning team. Second, the \texttt{git\_blame} node carries a postcondition that refuses to emit a draft when blame ranking returns no candidate commit. A free-form agent, lacking this guard, would previously fabricate a plausible cause from the stack trace alone. The routing primitive is equally deliberate about its own limits: \texttt{ANDROID\_NATIVE\_CRASH} is registered with an empty interceptor list, so the engine surfaces a localized, named ``no interceptor found'' outcome at the routing node rather than silently skipping the cluster or letting a downstream LLM hallucinate over an empty stack.

\paragraph{Outcome}
On a representative \texttt{Null\allowbreak Pointer\allowbreak Exception} regression in the feed adapter, the chain resolved the crash to a specific file and line, matched a historical case at score 0.83, attributed the regression to a recent switch to an asynchronous deletion path, and produced a remediation hint with a suggested owning team. Ingest-to-draft time was about eight minutes, against a pre-deployment mean of roughly four hours of manual triage. Draft tickets are routed to the correct owning team on the first try in about 85\% of cases, and the on-call engineer accepted this draft after a 100-second review.

\subsection{Cross-Cutting Observations}
\label{subsec:case_obs}

\begin{table}[t]
\centering
\caption{Production deployment outcomes for the two incident-diagnosis blueprints. Figures are rounded to one significant figure for anonymity; MTTR is alert/ingest to reviewed draft ticket.}
\label{tab:deployment}
\begin{tabular}{lcc}
\toprule
\textbf{Metric} & \textbf{Java OOM} & \textbf{Mobile Crash} \\
\midrule
Throughput & $\sim$40 clusters/wk & $\sim$800 clusters/day \\
MTTR before & $\sim$30 min & $\sim$4 h \\
MTTR after & $\sim$5 min & $\sim$8 min \\
LLM calls / incident & 2 & 2 \\
Unsafe-action incidents & 6/qtr $\rightarrow$ 0 & --- \\
False-positive rate & --- & 30\% $\rightarrow$ 8\% \\
Draft accepted as-is & $\sim$60\% & --- \\
Correct-team routing & --- & $\sim$85\% \\
\bottomrule
\end{tabular}
\end{table}

Two observations recur across both deployments and connect back to the benchmark results. First, the operationally decisive gains come not from the LLM reasoning better but from removing decisions from its responsibility: the unsafe-action count and false-positive rate both fell because a pre- or postcondition refused to let the workflow proceed on an unsafe or ungrounded state---the same mechanism that eliminates commonsense-constraint violations on TravelPlanner. The LLM is confined to suggestion and never performs a side effect; every draft ticket is human-reviewed before it is acted on.

Second, the framework changes how failures present themselves. When a structural assumption breaks---a heap dump with no usable suspects, a stack frame absent from the repository index, a crash type with no registered interceptor---the engine halts at a named node with a localized violation that points an engineer directly at the gap. Removing individual guards in controlled runs confirms this: without the OOM \texttt{correlate} postcondition the workflow root-caused an empty payload and emitted a fabricated leak hypothesis, and without the \texttt{git\_blame} postcondition the mobile chain targeted a nonexistent file---the same confidently-wrong output a free-form agent produced. The deterministic structure does not make the model more accurate; it makes failures debuggable, the property an on-call engineer relies on.

Table~\ref{tab:deployment} summarises the outcomes. These are operational figures from internal deployments rather than controlled experiments, and the two workflows differ in scale and maturity; dashes mark metrics not meaningful for a given workflow (e.g., the mobile pipeline emits draft tickets only, with no equivalent of the OOM unsafe-action hazard). We report them not as a benchmark but as evidence that the architecture transfers from a public planning benchmark to the procedurally well-defined, constraint-intensive SE workflows it is designed for.

\section{Discussion}
\label{sec:discussion}

\subsection{Limitations and Applicability}

While the Source Code Agent framework demonstrates significant improvements in constraint adherence and execution reliability, we discuss its design considerations and applicability scope.

\textbf{Blueprint Creation Process.} Creating execution blueprints requires domain expertise in both operational procedures and programming, but our framework lowers this barrier in three ways. The modular Controller, Executor, and Runtime Environment expose well-defined APIs, and the SDK's atomic, composite, and high-level tools let authors focus on domain logic (Section~\ref{sec:solution}). Specialized agents assist authoring---suggesting workflow structures, flagging constraint-validation points, and generating code templates from natural-language descriptions---so the LLM contributes creative synthesis while the engine preserves deterministic execution. Finally, a blueprint is reusable: once written for a domain it serves thousands of tasks at consistent reliability (35.56\% vs 18.00\% for prompt-based methods), amortizing the upfront cost.

To ground this cost rather than assert it, we report the authoring effort for the blueprints used in this work. The artifacts are compact: the HR leave-request workflow is 87 lines of code, the Android crash-triage blueprint 290, and the Java OOM-diagnosis blueprint 332, all written against the same engine API. Authoring is also fast in practice. Across 53 internally tracked workflows produced through the human--agent collaboration described above, more than 90\% converged to a passing blueprint---one that compiles and satisfies every node's pre- and postcondition on a smoke task with reviewer sign-off---within five human--agent dialogue rounds. The residual long-tail cases were not engine failures but specification problems: workflows that required a new control-flow primitive, or business rules that were self-contradictory and had to be reconciled with their owner before any blueprint could express them. This is consistent with our central claim, since the cost of authoring is paid once per workflow and the resulting artifact executes deterministically thereafter.

\textbf{Target Application Domains.} The framework targets structured, procedural tasks where reliability and constraint adherence are paramount: crash-diagnosis systems following established protocols, regulatory-compliance workflows with audit requirements, and other safety-critical operational procedures. In crash diagnosis, the blueprint codifies expert knowledge (check logs $\rightarrow$ analyze stack traces $\rightarrow$ query knowledge base $\rightarrow$ generate report) into a verifiable workflow junior engineers can rely on. For open-ended or rapidly evolving ad-hoc problems, general-purpose LLM agents remain more suitable; our contribution is reliable automation for the large class of operational tasks where stability matters more than adaptability.

\subsection{Broader Impacts}

\textbf{Reliability, Auditability, and Efficiency.} By eliminating workflow-level uncertainty through deterministic code, the framework lets agents operate where predictability and auditability are requirements: finance, healthcare, and safety-critical operations demand verifiable behavior, and the 96.0\% reduction in constraint violations shows code-based enforcement reaches regulated-domain reliability. Blueprints double as human-readable specifications auditors can inspect to verify compliance, and when a run fails, deterministic execution localizes the cause through code inspection rather than trajectory sampling---the accountability property black-box reasoning lacks. The 27.1\% step reduction over ReAct further lowers compute per task and makes execution time predictable enough for capacity planning.

\section{Conclusion}
\label{sec:conclusion}
We introduced and validated the \textsc{Source Code Agent} framework, which resolves LLM-agent non-determinism by codifying operational logic into deterministic source-code blueprints that constrain the LLM to a specialized tool within a predictable workflow rather than an unpredictable decision-maker. On TravelPlanner it achieves a 35.56\% final pass rate (97.6\% over ATLAS on the same backbone), a 96.0\% reduction in constraint violations (11 vs 275), and a 27.1\% efficiency gain (10.2$\pm$0.7 vs 14.0 steps). A controlled ablation attributes these gains to deterministic enforcement rather than richer specification, and production deployments plus ScienceWorld/ALFWorld results show the architecture transfers beyond a single benchmark. The framework's payoff scales with how procedurally well-defined and constraint-intensive a workflow is; general-purpose agents remain better for open-ended tasks. It is a step toward verifiable, auditable autonomy, and motivates automating blueprint authoring.

\bibliographystyle{IEEEtran} 
\bibliography{ref} 

\end{document}